\documentclass[useAMS,usenatbib]{mn2e}
\usepackage{graphicx}
\usepackage{float}
\usepackage{wrapfig}
\usepackage{booktabs}

\title[Tully Fisher evolution]{Size matters: abundance matching,  galaxy sizes, and the Tully-Fisher relation in EAGLE}

\author[Ferrero et al.]{
\parbox[t]{\textwidth}{
Ismael Ferrero$^{1,2}$\thanks{E-mail: iferrero@oac.unc.edu.ar},
Julio F. Navarro$^{3,4}$,
Mario G. Abadi$^{1,2}$,
Laura V. Sales$^{5}$,\\
Richard G. Bower$^{6}$,
Robert A. Crain$^{7}$,
Carlos S. Frenk$^{6}$,
Matthieu Schaller$^{6}$,\\
Joop Schaye$^{8}$ and
Tom Theuns$^{6}$
}
\\
\\
$^{1}$Instituto de Astronom\'ia Te\'orica y Experimental, CONICET-UNC, Laprida 854, X5000BGR, C\'ordoba, Argentina\\
$^{2}$Observatorio Astron\'omico de C\'ordoba, Universidad Nacional de C\'ordoba, Laprida 854, X5000BGR, C\'ordoba, Argentina\\ 
$^{3}$Department of Physics \& Astronomy, University of Victoria, Victoria, BC V8P 5C2, Canada\\
$^{4}$Senior CIfAR Fellow\\
$^{5}$Department of Physics and Astronomy, University of California, Riverside, CA, 92521, USA\\
$^{6}$Institute for Computational Cosmology, Department of Physics, Durham University, South Road, Durham, DH1 3LE, UK\\
$^{7}$Astrophysics Research Institute, Liverpool John Moores University, 146 Brownlow Hill, Liverpool L3 5RF, UK\\
$^{8}$Leiden Observatory, Leiden University, P.O. Box 9513, 2300 RA Leiden, the Netherlands\\}

\begin{document}

\maketitle 
\date{}
\pubyear{0000}
\maketitle
\begin{abstract}
  The Tully-Fisher relation (TFR) links the stellar mass of a disk
  galaxy, $M_{\rm str}$, to its rotation speed: it is well
  approximated by a power law, shows little scatter, and evolves
  weakly with redshift. The relation has been interpreted as
  reflecting the mass-velocity scaling ($M\propto V^3$) of dark matter
  halos, but this interpretation has been called into question by
  abundance-matching (AM) models, which predict the galaxy-halo mass
  relation to be non-monotonic and rapidy evolving. We study the TFR
  of luminous spirals and its relation to AM using the EAGLE set of
  $\Lambda$CDM cosmological simulations. Matching both relations
  requires disk sizes to satisfy constraints given by the
  concentration of halos and their response to galaxy assembly. EAGLE
  galaxies approximately match these constraints and show a tight
  mass-velocity scaling that compares favourably with the observed
  TFR.  The TFR is degenerate to changes in galaxy formation
  efficiency and the mass-size relation; simulations that fail to
  match the galaxy stellar mass function may fit the observed TFR if
  galaxies follow a different mass-size relation.  The small scatter
  in the simulated TFR results because, at fixed halo mass, galaxy
  mass and rotation speed correlate strongly, scattering galaxies
  along the main relation. EAGLE galaxies evolve with lookback time
  following approximately the prescriptions of AM models and the
  observed mass-size relation of bright spirals, leading to a weak TFR
  evolution consistent with observation out to $z=1$. $\Lambda$CDM
  models that match both the abundance and size of galaxies as a
  function of stellar mass have no difficulty reproducing the observed
  TFR and its evolution.
\end{abstract}

\section{Introduction}
\label{SecIntro}

The Tully-Fisher relation (TFR) links the luminosity of disk galaxies
with their characteristic rotation speed. First noted by
\citet{Tully-Fisher1977} using photographic magnitudes and HI velocity
widths, it has become one of the best studied galaxy scaling relations
and a powerful secondary distance indicator. It is well approximated,
for luminous spirals, by a tight power law whose dependence on
wavelength is fairly well understood
\citep{Aaronson1979,Mathewson-Ford-Buchhorn1992,Verheijen1997,Tully1998,Haynes1999,Courteau2007}. As
a result, the relation is now routinely cast in terms of galaxy
stellar mass and the circular speed measured at a characteristic
``luminous radius'' \citep{Bell2001,Reyes2011,Avila-Reese2008}. Since
rotation curves are nearly flat the choice of radius is not critical
for luminous spirals, but popular choices include $2.2$ times the
exponential scalelength \citep[e.g.,][]{Courteau1997} or,
alternatively, a radius that contains roughly $80\%$ of all stars
\citep[e.g.,][]{Pizagno2007}.

The evolution of the TFR with redshift has been more difficult to pin
down, although the consensus is that the TFR evolves weakly, if
at all, up to $z \approx 1$.  Early studies, many of them in the B-band,
claimed significant evolution in either the zero-point, the slope or
in both \citep[e.g.][]{Ziegler2002,Bohm2004}, but these conclusions
evolved once data on longer wavelengths less affected by extinction
became available.  \citet{Conselice2005} and \citet{Flores2006}, for
example, found no significant evolution in the K-band TFR to $z \approx 1.3$
and $z \approx 0.6$, respectively.  This conclusion has been supported by
the more recent work of \citet{Miller2011}, who conclude that there is
no substantial change in the stellar-mass TFR out to redshifts of
about unity.  Observations at higher redshifts hint at more
substantial evolution of the zero-point although the presence of large
random motions and selection effects at such early times complicate
the interpretation
\citep{Forster-Schreiber2009,Cresci2009,Kassin2012}.

The properties of the TFR have long challenged direct numerical
simulations of disk galaxy formation in the $\Lambda$CDM scenario.
Early work, for example, produced galaxies so massive and compact that
their rotation curves were steeply declining and, at given galaxy
mass, peaked at much higher velocities than observed \citep[see,
e.g.,][and references
therein]{Navarro2000,Abadi2003,Scannapieco2012}. The problem was
quickly traced to the inability of early feedback schemes to prevent
large amounts of low-angular momentum baryons from accumulating early
at the center of dark matter halos.

Subsequent work made progress by adopting feedback schemes able to
remove a large fraction of the early-collapsing baryons and to
regulate their further accretion, leading to disks with sizes and
rotation curves in better accord with observation
\citep[e.g.,][]{Okamoto2005,Governato2007,Brook2011,McCarthy2012,Guedes2013,Aumer2013,Marinacci2014}.
Although such results were promising, they were inconclusive,
especially because they were either based on a handful of carefully
selected, and therefore likely highly biased, individual systems, or
on cosmological boxes where simulated galaxies failed to match basic
statistics of the observed galaxy population, such as the galaxy
stellar mass function.

As a result, much theoretical TFR work in the context of the
$\Lambda$CDM cosmology has proceeded via semi-analytic models of
galaxy formation. These models employ simple, albeit well-founded,
prescriptions to generate a synthetic galaxy population within an evolving
population of dark matter halos. The physical properties of such a
population are then compared with observed galaxies in order to
calibrate the assumed prescriptions and to shed light onto the role of
various mechanisms during galaxy formation \citep[see, e.g.,][and
references therein]{Cole2000,Dutton2010}.  Semi-analytic models have
highlighted a number of difficulties, particularly when attempting to
match simultaneously the abundance of galaxies as a function of
stellar mass and the slope and normalization of the TFR
\citep[see][for recent attempts]{Lacey2015,Desmond2015}.

The basic reason for these difficulties is that these models generally
(and reasonably) assign more massive galaxies to more massive halos,
leading to a tight relation between galaxy and halo masses that places
strong constraints on their characteristic circular speed.  A simple
model for this galaxy-halo mass relation may be derived by ranking
galaxies by mass and assigning them to halos ranked in similar
fashion, preserving the ranked order
\citep{Frenk1988,Vale-Ostriker2004,Guo2010,Moster2013,Behroozi2013}.  This
``abundance-matching'' (AM) exercise has proven particularly useful
when assessing the results of numerical simulations, especially those
of single isolated systems, where there is otherwise little guidance
about the mass or size of the galaxy that may form in one particular
halo.

Since the dark mass profile of $\Lambda$CDM halos is well known
\citep{Navarro1996,Navarro1997}, AM models have little freedom left
when trying to match the TFR: a galaxy's characteristic circular
velocity is fixed once its radius and the halo response have been
specified \citep[see, e.g.,][]{Cattaneo2014}.  The critical role of
galaxy size and halo response implies that insight into the origin of
the TFR requires a good understanding of the interplay between
galaxies' mass and size, as well as of the mass of the halos they
inhabit and how galaxies might modify them.

These complex issues are best studied through cosmological
hydrodynamical simulations, especially those able to follow
statistically significant numbers of galaxies over large volumes, and to
resolve their inner regions, where rotation speeds are
measured.  These conditions are well met by the latest round of
cosmological hydrodynamical simulations, such as the
recently-completed Illustris and EAGLE projects
\citep{Vogelsberger2013,Schaye2015}. One main conclusion from these
efforts is that, except for the lowest masses \citep{Sawala2013,Sawala2015},
abundance-matching predictions are actually quite 
robust: matching the observed galaxy stellar mass function requires
simulations to reproduce accurately the galaxy-halo mass relation
envisioned by AM models, with little scatter. 


One intriguing result, however, is that both Illustris and EAGLE
report good agreement with the observed TFR, despite the fact that the
galaxy stellar mass functions they report differ significantly. This
approximate ``invariance'' of the simulated TFR has been noted in the
past. \citet{Guo2010}, for example, found that a number of simulated
galaxies, which in earlier work had been reported to match the TFR,
actually had masses that greatly exceeded AM predictions.  A similar
result has been discussed recently by \citet{Torrey2014}, who showed
that the TFR in their simulations is insensitive to large variations
in the Illustris galaxy formation physics submodules: only models with
``no feedback'' were found to be in substantial disagreement with the
observed TFR. Although \citet{Torrey2014} cite ``feedback'' as an
essential ingredient to match the TFR, its actual role in determining
its slope and zero-point remains unclear, a point underlined by the
recent results of \citet{Crain2015}, who report that the TFR is
actually quite sensitive to feedback, at least in their
implementation.

We examine these issues here using the EAGLE set of $\Lambda$CDM
cosmological hydrodynamical simulations. We analyze the stellar mass
TFR in the regime of luminous spirals, where gas contributes little to
the overall baryon budget, and report results on the baryonic Tully-Fisher
relation of gas-dominated, fainter galaxies in a separate paper
\citep{Sales2016}. We pay particular attention to the effect of galaxy
sizes on the TFR, an issue that has been relatively well explored in
semi-analytic approaches but that has received little attention in
direct simulation TFR work. 

We begin in Sec.~\ref{SecModel} by motivating the effect of galaxy size
on the TFR by simple considerations that highlight the need for halo
contraction in order to reconcile the TFR with the predictions of
abundance matching models. We then present, in Sec.~\ref{SecResults}, 
the TFR of simulated galaxies in EAGLE, with particular attention to
the origin of its small scatter (Sec.~\ref{SecScatter}) and its
evolution (Sec.~\ref{SecTFEVol}). We conclude with a
brief summary of our main findings in Sec.~\ref{SecConc}.

\begin{center} 
\begin{figure*}
\advance\leftskip-0.85cm
\includegraphics[width=200mm]{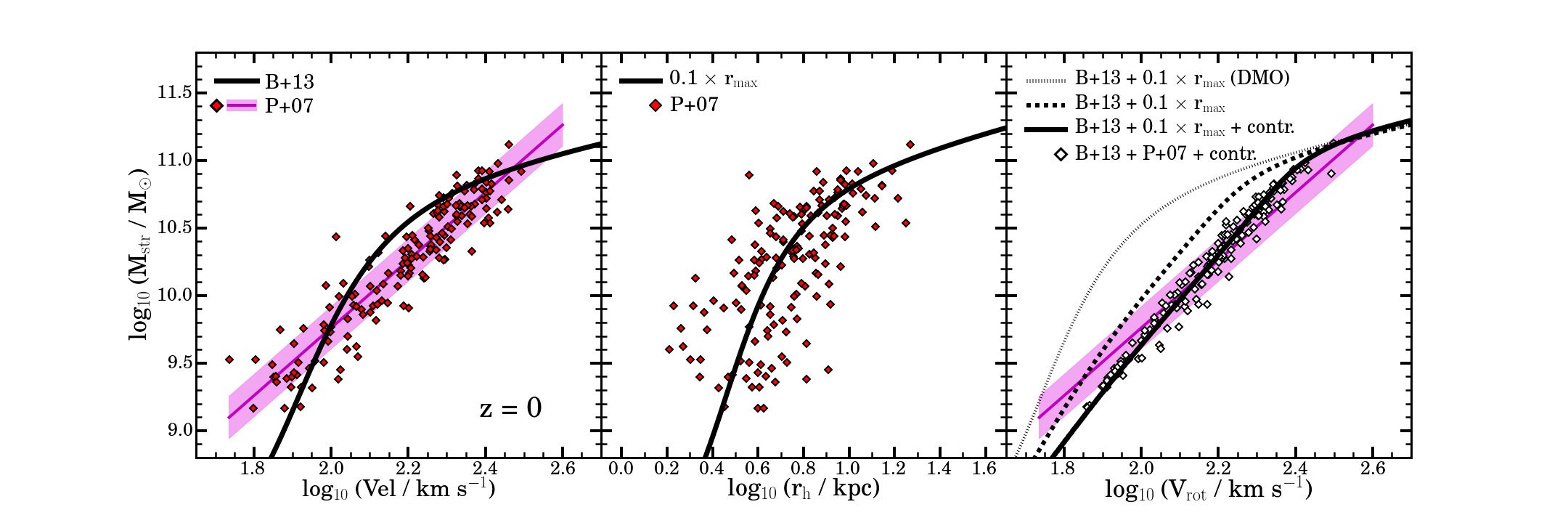}
\caption{Galaxy stellar mass, $M_{\rm str}$, as a function of various
  parameters. {\it Left:} The solid black curve shows the
  abundance-matching prediction of \citet[][B+13]{Behroozi2013}, as a
  function of halo virial velocity, $V_{200}$. Symbols correspond to
  the data of \citet[][P+07]{Pizagno2007}, converted to stellar masses
  using a constant I-band mass-to-light ratio of $1.2$
  \citep[][]{Bell2003} and shown as a function of disk rotation speed,
  $V_{\rm rot}$. Color-shaded band indicates the mean slope and
  $1$-$\sigma$ scatter. {\it Middle:} Symbols show half-light radii of
  galaxies in the P+07 sample. Thick solid line indicates a multiple
  of $r_{\rm max}$, the characteristic radius where NFW halo circular
  velocities peak. Halo masses are as in the B+13 model of
  the left panel.  {\it Right:} Tully-Fisher relation. The color band
  is the same as in the left-hand panel. The dotted curve indicates
  the dark halo circular velocity at $r_h=0.1\, r_{\rm max}$, assuming
  NFW profiles and neglecting the contribution of the disk. The dashed
  line includes the gravitational contribution of the disk, keeping
  the halo unchanged. Finally the thick solid line (and symbols)
  include the disk contribution {\it and} assume that halos contract
  adiabatically.}
\label{FigModel}
\end{figure*}
\end{center}

\begin{center} 
\begin{figure*}
\advance\leftskip-0.1cm
\includegraphics[width=180mm]{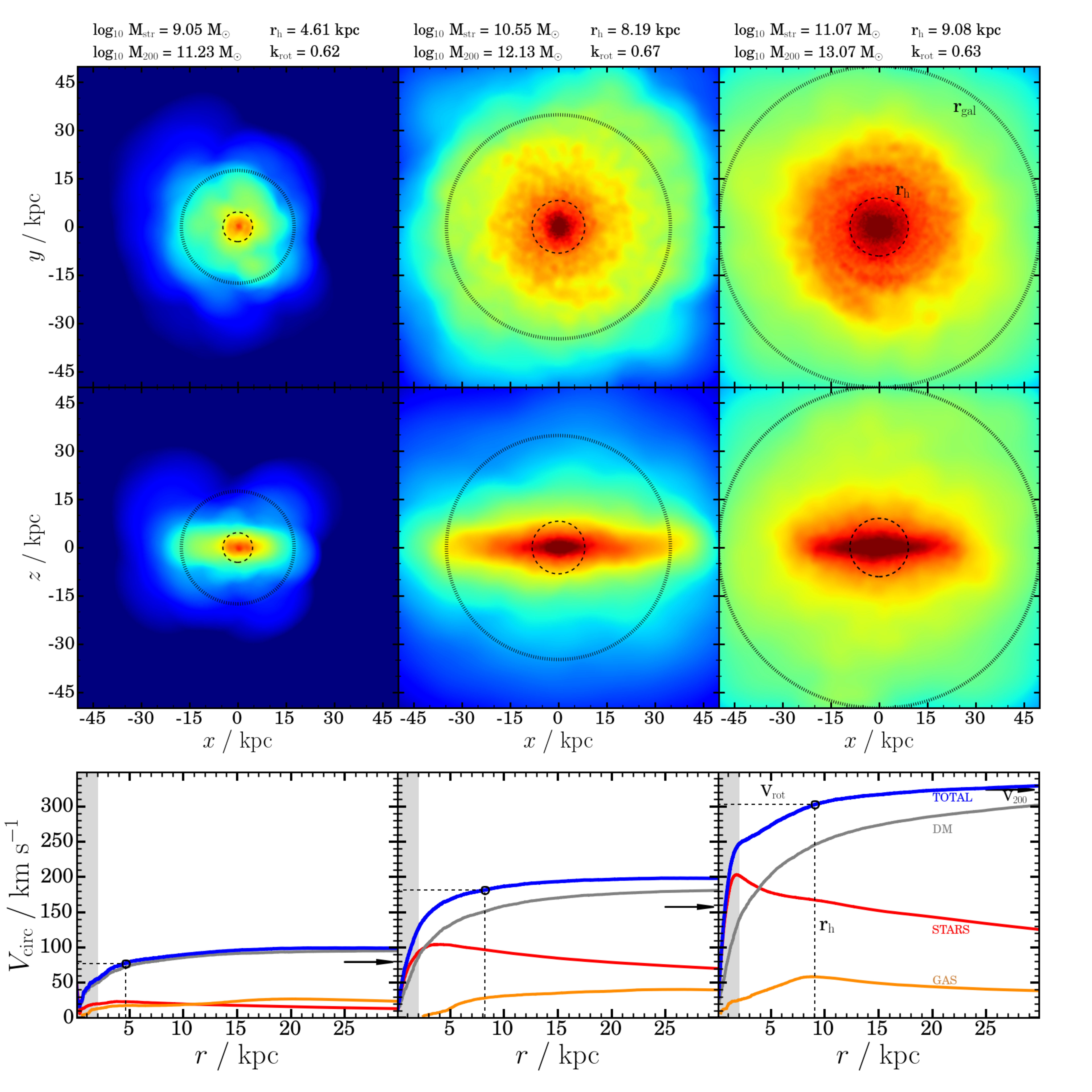}
\caption{Stellar surface density maps of three simulated disk galaxies
  at $z=0$.  Stellar and halo masses, half-mass radii, and rotation
  parameter values are listed in the legends. The top row show a
  face-on view of the disks, the middle rows show edge-on views. The
  inner and outer circles indicate stellar half-mass radii, $r_h$, and
  $r_{\rm gal}=0.15\, r_{200}$, respectively. The corresponding
  circular velocity curves are shown in the bottom row. Blue denotes
  total circular velocity, grey the dark matter contribution, red the
  stars and orange the gas. Stellar half-mass radii and rotation
  speeds, $V_{\rm rot}=V_{\rm circ}(r_h)$, are indicated by dotted
  lines. The halo virial velocity, $V_{200}$, is shown with a
  horizontal arrow in each bottom panel. }
\label{FigImages}
\end{figure*}
\end{center}

\begin{center} 
\begin{figure*}
\advance\leftskip-0.9cm
\includegraphics[width=200mm]{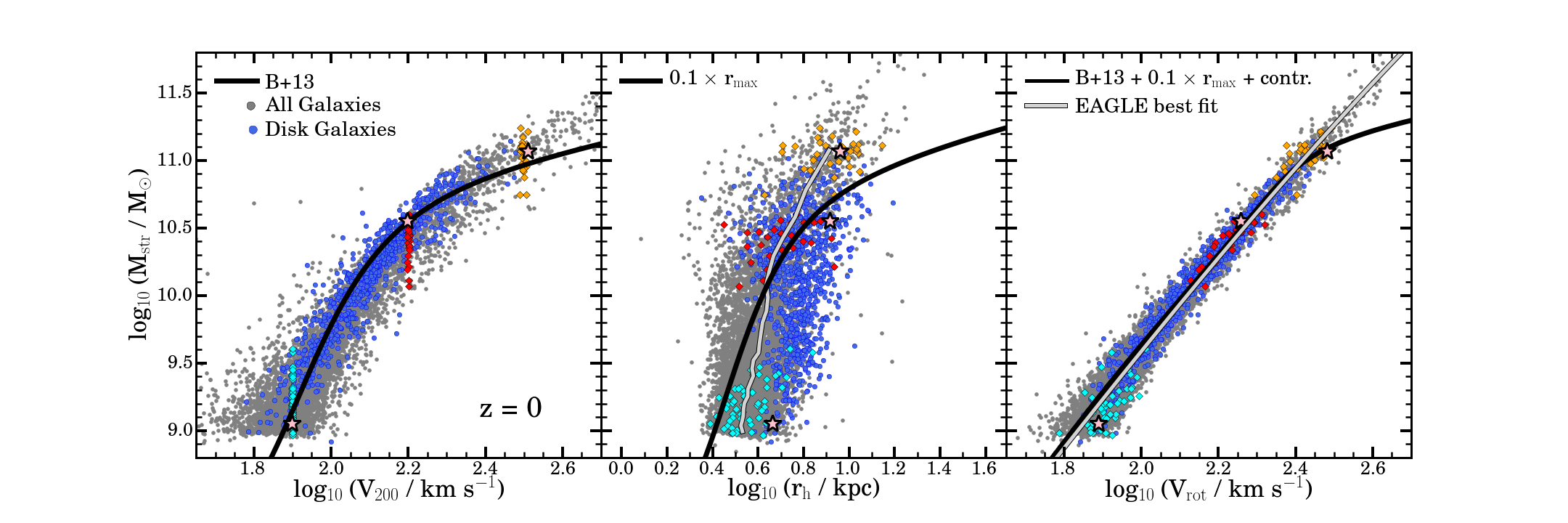}
\caption{Analogous to Fig.~\ref{FigModel}, but 
  for EAGLE galaxies at $z=0$. Black solid lines are as in
  Fig.~\ref{FigModel}, and are included to aid comparison. Grey points
  correspond to all simulated galaxies, blue points indicate 
  ``disks'' according to a relatively strict criteria; i.e., systems
  with rotation parameter $\kappa_{\rm rot}>0.6$. Galaxies
  forming in three narrow bins of halo mass are highlighted in cyan, red,
  and orange. The three starred symbols indicate the three
  galaxies shown in the images of Fig.~\ref{FigImages}. Note that
  EAGLE galaxies: (i) follow closely the B+13 abundance-matching
  predictions (left), (ii) have sizes comparable to spirals in the TF
  sample of P+07 (middle); and (iii) have a Tully-Fisher relation in
  good agreement with the predictions from the simple halo contraction
  model of Fig.~\ref{FigModel} (right). The thin grey line is a fit to
  the simulated TFR; see parameters in Table~\ref{TabFitParam}. }
\label{FigSimTFR}
\end{figure*}
\end{center}

\vspace*{-15mm}
\section{Tully-Fisher, abundance matching, and galaxy sizes}
\label{SecModel}

In a cosmological context, the Tully-Fisher relation has often been
thought to reflect the equivalence between halo mass and circular
velocity imposed by the finite age of the Universe \citep[see,
e.g.,][]{Mo1998,Steinmetz1999}. That characteristic timescale
translates into a fixed density contrast that implies a scaling
between virial\footnote{Virial quantities are identified by a ``200''
  subscript and measured at the virial radius, $r_{200}$, defined as
  the radius where the enclosed mean density is $200$ times the
  critical density of the Universe, 
  $\rho_{\rm crit}=3H^2(z)/8\pi G$.} mass and circular velocity
given by
\begin{equation}
\label{EqM200V200}
M_{200} = \frac{V_{200}^3}{10 G H(z)},
\end{equation}
where $G$ is the gravitational constant and $H(z)$ is the Hubble
constant. The power-law scaling resembles the TFR, provided that
galaxy masses and rotation velocities scale roughly in proportion to the
virial masses and circular velocities of the halos they inhabit
\citep[][]{Navarro2000}. 

This interpretation, however, is difficult to reconcile with
abundance-matching arguments, which suggest that galaxy masses are not
simply proportional to the virial mass of the halo they
inhabit. Indeed, AM models require the ``galaxy formation
efficiency'', $f_m \equiv M_{\rm str}/M_{200}$, to be a non-monotonic function
of virial mass, reaching a maximum at
$M_{200}\sim 10^{12} \, $M$_{\odot}$ and declining steeply toward lower
and higher masses \citep[e.g.,][]{Eke2005,Guo2010,Moster2013,Behroozi2013}. In
addition, galaxy formation efficiencies evolve relatively rapidly 
in AM models, in contrast with the weak evolution of the TFR discussed
in Sec.~\ref{SecIntro}. A simple proportionality between the mass and
velocity of galaxies and halos is thus clearly inadequate to explain the
TFR within the context of AM models.

We show this in the left panel of Fig.~\ref{FigModel}, where the solid
black curve indicates the relation between $M_{\rm str}$, the stellar
mass of a galaxy, and $V_{200}$, the circular velocity of its halo,
according to the model of \citet[][hereafter, B+13]{Behroozi2013}. For
comparison, the red symbols show the Tully-Fisher data from the sample
of \citet[][hereafter, P+07]{Pizagno2007}. This comparison shows that
reconciling AM predictions with the TFR requires the relation between
disk rotation speed, $V_{\rm rot}$, and halo virial velocity to be
non-monotonic: rotation speeds must underestimate or overestimate halo
circular velocities, depending on $M_{\rm str}$. In other words, if
$f_m$ is non-monotonic, then the ratio
$f_v \equiv V_{\rm rot}/V_{200}$ cannot be monotonic either
\citep[see][for a similar analysis]{Cattaneo2014}.

Rotation speeds are set by the total (dark plus luminous) mass
enclosed within a given radius, so the ratio $f_v$ depends
sensitively on galaxy size: in general, at given $M_{\rm str}$, the
smaller the galaxy the larger the contribution of the disk to $V_{\rm
  rot}$ and the lesser the importance of the dark
halo. Quantitatively, the result depends on the density profile of the
halo and on its response to the assembly of the galaxy. 

It is instructive to consider each effect separately. We begin by
noting that the half-light radii, $r_h$, of galaxies in the P+07
sample (shown in the middle panel of Fig.~\ref{FigModel}) are well
approximated by a simple multiple of $r_{\rm max}$, the characteristic
radius where $\Lambda$CDM halos reach their peak circular velocity,
$V_{\rm max}$. This means that, if the luminous component was
unimportant, disk rotation speeds, identified with the circular
velocity at $r_h$, would be just a multiple of the virial
velocity. This is shown by the dotted curve in the right-hand panel of
Fig.~\ref{FigModel}, where we have assumed that halos follow an NFW
profile \citep{Navarro1996,Navarro1997} with concentrations given by
the $M_{200}$-$c$ relation derived from the latest set of large
cosmological N-body simulations \citep[][their Appendix
C]{Ludlow2016}. Clearly, this provides a rather poor match to the
observed TFR.

Simply adding in quadrature the velocity contribution of the luminous
galaxy (keeping the halos unchanged) tilts and shifts the relation to
that indicated by the thick dashed curve in the same panel. The tilt
in slope results because the gravitational importance of the disk
(within $r_h$) increases with increasing $M_{\rm str}$. The predicted rotation
velocities are, however, still below observed values, suggesting that,
if AM predictions hold, halo contraction is needed to explain the
observed TFR.

Modeling the halo response as ``adiabatic contraction''
\citep{Barnes1984,Blumenthal1986} yields the individual symbols and
the thick solid curve in the right-hand panel of
Fig.~\ref{FigModel}. Contraction compounds the gravitational effect of
the disk, tilting and shifting the relation further, and leading to a
reasonably good agreement with the observed TFR, despite the
simplicity of the model and the fact that we have not allowed for any
adjustment or scatter in the AM prediction or in the
mass-concentration relation. Note that the curvature in the
mass-velocity relation characteristic of AM models is shifted to large
velocities, where there is little data, resulting in a TFR that can be
adequately approximated by a single power law with little scatter.

These results suggest that $\Lambda$CDM models should be able to match
the observed TFR, {\it provided} that galaxy sizes are well
reproduced, and that halos respond roughly adiabatically to galaxy
assembly. In particular, much smaller galaxy sizes would lead to
excessively high rotation velocities at given galaxy mass, the main
reason for the failure of early disk galaxy formation simulations
discussed in Sec.~\ref{SecIntro}. This discussion also illustrates
that the TFR is a sensitive probe not only of the galaxy-halo mass
relation, but also of galaxy sizes: indeed, models that deviate from
the AM predictions may still fit the TFR if galaxy sizes are adjusted
appropriately. We explore next whether these simple insights hold when
analyzing the TFR and its redshift evolution in a large cosmological
hydrodynamical simulation.

\section{Numerical results}
\label{SecResults}

\subsection{The EAGLE simulation}
\label{SecEAGLE}
 
We use here the EAGLE\footnote{http://eagle.strw.leidenuniv.nl} set of
cosmological hydrodynamical simulations. We briefly summarize the
main relevant aspects of these simulations and refer the interested
reader to \citet{Schaye2015} for further details.

EAGLE used a heavily modified version of GADGET-3, a itself-modified
version of the Tree-SPH hydrodynamical code GADGET-2 \citep{Springel2005}.
The modifications to the hydrodynamics solver are described and their 
effects investigated in \citet{Schaller2015a}.The simulation includes subgrid models 
for radiative cooling \citep{Wiersma2009a}, star formation \citep{Schaye-DallaVecchia2008},
 stellar mass loss \citep{Wiersma2009b}, energetic feedback from star formation
 \citep{DallaVecchia-Schaye2012}, black hole accretion, merging and feedback 
 \citep{Springel2005, Schaye2015,Rosas-Guevara2015}. The simulation assumes a 
$\Lambda$ Cold Dark Matter cosmology with parameters consistent with the 
latest CMB experiments \citep{PlanckCollaboration2014}: $\Omega_b = 0.0482$,
$\Omega_{dark} = 0.2588$, $\Omega_{\Lambda} = 0.693$, and $h = 0.6777$,
where $H_0 = 100 \, h\, \rm km\ s^{-1} Mpc^{-1}$.

The simulation analyzed here is referred to as Ref-L100N1504 in Table
1 of \citet{Schaye2015}. It follows $2 \times 1504^3$ particles in a
periodic cubic volume of $100$ Mpc on a side from redshift $z=127$ to
$z=0$. This corresponds to an equal number of gas and dark matter
particles with initial mass of
$m_{gas}=1.81 \times 10^6\; \rm M_{\odot}$ and
$m_{DM}=9.70 \times 10^6\; \rm M_{\odot}$ per particle. The simulation
uses a Plummer-equivalent gravitational softening of
$\epsilon = 2.66 $ kpc (comoving units) before redshift $z=2.8$ and
fixed at $\epsilon=0.7$ kpc (physical units) after that.  The
numerical parameters in the {\sc eagle} subgrid physics modules for feedback
have been calibrated to the observed $z = 0$ galaxy stellar
mass function and the distribution of galaxy sizes.

\citet{Schaye2015} have presented a preliminary version of the
TFR, using the maximum circular velocity as a proxy for the disk
rotation speeds. \citet{Crain2015} have also analyzed the TFR,
especially its dependence on feedback strength; they report an
increase in rotation speed at fixed mass when feedback is more
efficient (see their Fig. 10d). Our analysis extends this work by
focussing on velocities measured at the half-mass radii of the
simulated galaxies (closer to what is actually observed), by
considering the importance of galaxy sizes, and by examining the
evolution of the TFR with redshift.

\subsection{The simulated galaxy sample}
\label{SecGxSample}

Galaxies are identified in EAGLE using the SUBFIND algorithm
\citep{Springel2001,Dolag2009}, which selects gravitationally bound
substructures (subhalos) in halos found by a friends-of-friends (FoF)
algorithm with linking length 0.2 times the mean interparticle
separation \citep{Davis1985}. The centre of each subhalo is defined as
the position of the member particle with the minimum gravitational
potential energy. Galaxy properties are computed within a ``galaxy
radius'' defined by the smaller of either $r_{\rm gal}=0.15 \ r_{200}$
or $50$ kpc, a choice that encompasses most of the stars in each halo, as well
as the majority of its cold gas.

We focus here on ``central'' galaxies (i.e., those corresponding to
the most massive subhalo in each FoF group) with a minimum stellar
mass of $M_{\rm str}=10^9\, $M$_{\odot}$ (i.e., about $700$ star
particles). We shall show results for all galaxies, as well as for
``disk'' galaxies, defined as those whose rotational-to-total kinetic
energy parameter $\kappa_{\rm rot} = \Sigma V_{xy}^2/\Sigma V^2 > 0.6$
\citep{Sales2012}. (Here $V$ is the magnitude of the total velocity
vector and $V_{xy}=j_z/R$ its azimuthal component perpendicular to the
$z$-direction, which is defined by the total angular momentum of the
galaxy's stellar component.) Note that this criterion is quite
strict, and selects only $11\%$ of all galaxies as disks at $z=0$. Our final
galaxy samples contain $7482$ galaxies ($867$ of them disks) at $z=0$,
and $7922$ galaxies ($818$ of them disks) at $z=1$.

Fig.~\ref{FigImages} shows the spatial distribution of the stellar
component for three of our simulated disk galaxies, spanning a wide
range of $M_{\rm str}$, from $\sim 10^9$ to $10^{11}\,
$M$_{\odot}$.
The three galaxies are shown face-on (top) and edge-on (middle) and
have been chosen to have well-defined disks (i.e.,
$\kappa_{\rm rot}>0.6$).  Their circular velocity curves, here approximated by
$V_{\rm circ}(r)=(GM(<r)/r)^{1/2}$, are shown in the bottom row and
are approximately flat in the inner $10$-$30$ kpc
\citep{Schaller2015b}. Hereafter, we shall use the circular velocity at
the stellar half-mass radius (shown by a dotted vertical line or
circle in each panel) to define the characteristic rotation speed
associated with each galaxy. Although formally the circular velocity
of an axisymmetric disk will differ from the definition provided above 
(which is correct for spherical systems), the corrections are typically 
smaller than $10\%$. Therefore, for simplicity, we apply the same 
definition of circular velocity regardeless of galaxy morphological type.

\subsection{The simulated Tully-Fisher relation}
\label{SecSimTFR}

Fig.~\ref{FigSimTFR} summarizes the results of our simulation
regarding abundance matching, galaxy sizes and the TFR at $z=0$. The
black solid curves are reproduced from Fig.~\ref{FigModel}, for ease
of comparison. Individual simulated galaxies are shown in grey, disk
galaxies in blue. In addition, all central galaxies in three narrow bins of
halo mass are identified and highlighted in cyan, red, and orange to
guide the discussion.

The leftmost panel shows that EAGLE follows  the
results of the B+13 abundance matching model. The agreement
is not perfect, however, and leads
to slight but systematic deviations in the galaxy stellar mass
function which, around its knee, is offset from the observational inferred
relation by about a factor of two. Disks follow the main galaxy mass-halo mass
relation quite well, with a hint of higher-than-average galaxy
formation efficiencies at fixed halo mass.

Simulated disks are also slightly larger than spheroids at given
$M_{\rm str}$, as shown in the middle panel of Fig.~\ref{FigSimTFR}
\citep[see][]{Furlong2015b}. The overall mass-size trend of simulated
galaxies, however, is not far from that of the P+07 Tully-Fisher
sample, as indicated by the solid black line, which is the same as in
Fig.~\ref{FigModel}. There is, however, a slight mismatch, which
becomes evident at large masses, where EAGLE disks are smaller than in
the observed sample, and at small masses, where the opposite is true.

Finally, the right-hand panel of Fig.~\ref{FigSimTFR} shows the
simulated Tully-Fisher relation, and compares it with the
adiabatically-contracted model of Fig.~\ref{FigModel} (black solid
line). The grey solid curve shows a fit of the form\footnote{This is
  the same fitting form proposed by \citet{Sales2016} to describe the
  simulated ``baryonic Tully-Fisher relation'' of APOSTLE and EAGLE
  galaxies.}  $M_{\rm str}/{\rm M}_\odot=M_0 \, \nu^{\alpha}
\exp(-\nu^{\gamma})$, where $\nu$ is the velocity in units of $50$
km/s, $M_0=8.63\times10^{8}$, $\alpha=4.1$ and $\gamma=0.432$.  The
good agreement between the grey and black curves suggests that the
simple considerations discussed in Sec.~\ref{SecModel} capture the
basic ingredients of the relation between abundance matching, galaxy
sizes, and halo response seen in the EAGLE simulation.

Three points are worth emphasizing: (i) the TFR may be approximated by
a single power law, and is much straighter than the AM mass-velocity
relation; (ii) the TFR scatter is rather small, with an rms of $0.11$
dex in mass, or $0.08$ dex in velocity, and (iii) the TFR zero-point at
$M_{\rm str}=10^{10}\, $M$_{\odot}$ (roughly the mid-point of the mass
range considered here) is in excellent agreement with observation.

\begin{center} 
\begin{figure}    
\advance\leftskip-0.2cm
\includegraphics[width=93mm]{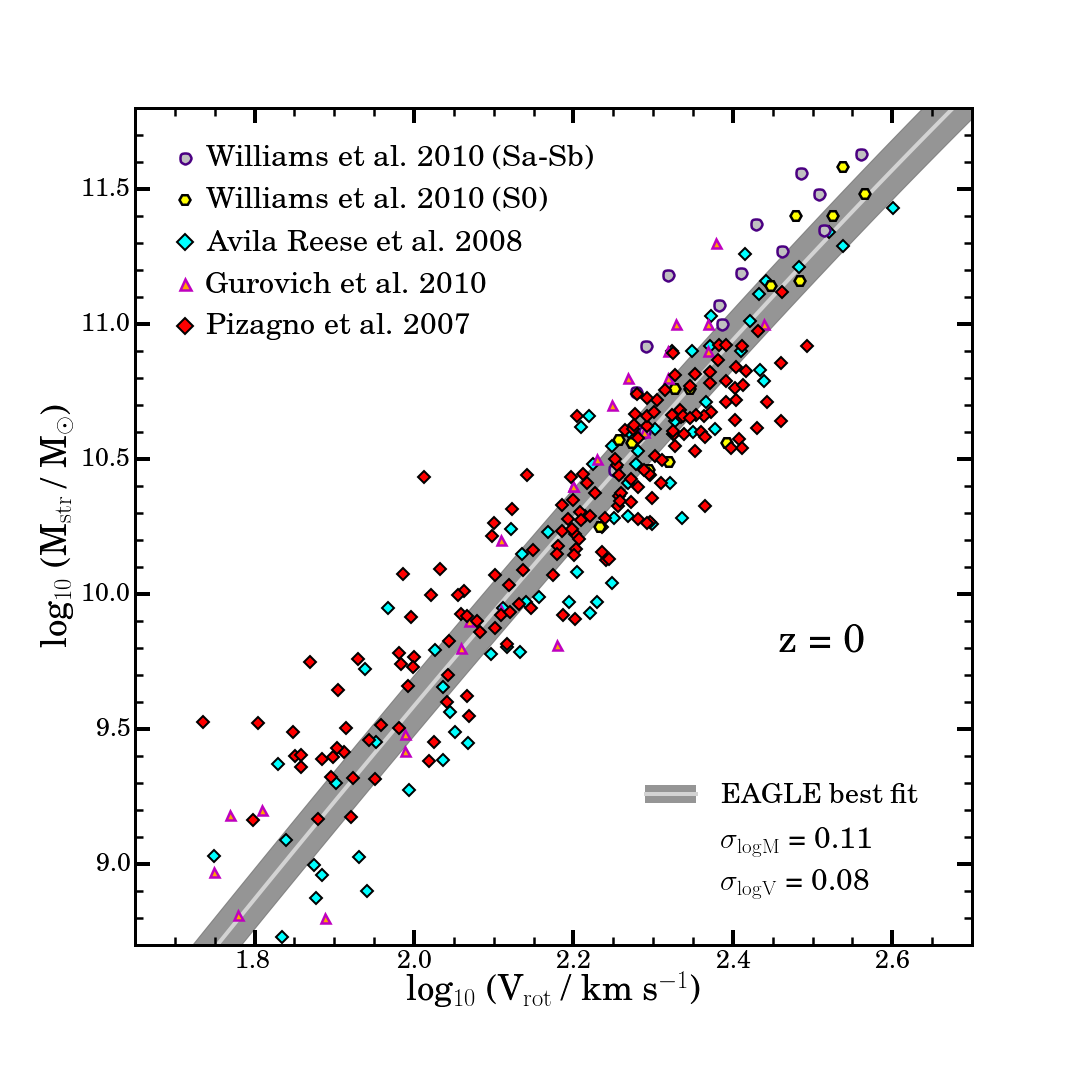} 
\vspace*{-10mm}
\caption{Tully-Fisher relation for EAGLE galaxies (grey band) compared
  with individual spirals taken from five recent TF compilations. 
  The simulated relation is in excellent agreement with the
  observational data. The scatter is even smaller than in observed
  samples, even though the simulated relation includes {\it all}
  galaxies and not only disks.}
\label{FigSimTFR2}
\end{figure}
\end{center}

A more direct comparison with observation is shown in
Fig.~\ref{FigSimTFR2}, where we plot the TFR for all EAGLE galaxies
(shown by a grey band to illustrate the main relation
$\pm 1$-$\sigma$) together with data from five recent Tully-Fisher
compilations\footnote{The observational data assume stellar masses
  derived assuming a Chabrier initial mass function (IMF). For
  galaxies in the \citet{Avila-Reese2008} compilation we have reduced
  their published stellar masses by 0.15 dex in order to convert from
  Salpeter to Chabrier IMF.}. EAGLE agree with these datasets quite
well, even when luminous early-type spirals from \citet{Williams2010}
are added to the comparison. Note that the scatter in the simulated
TFR is smaller than observed, even when considering {\it all
  galaxies}. Choosing only disks reduces the scatter even further, to
$\sim 0.09$ dex in mass and $0.07$ dex in velocity. We conclude that
the EAGLE TFR is in excellent agreement with observations at $z=0$. We
extend this analysis to higher redshifts in Sec.~\ref{SecTFEVol},
after exploring next what sets the slope, zero-point and scatter of
the simulated TFR.

\begin{center} 
\begin{figure*}    
\advance\leftskip-0.5cm
\includegraphics[width=155mm]{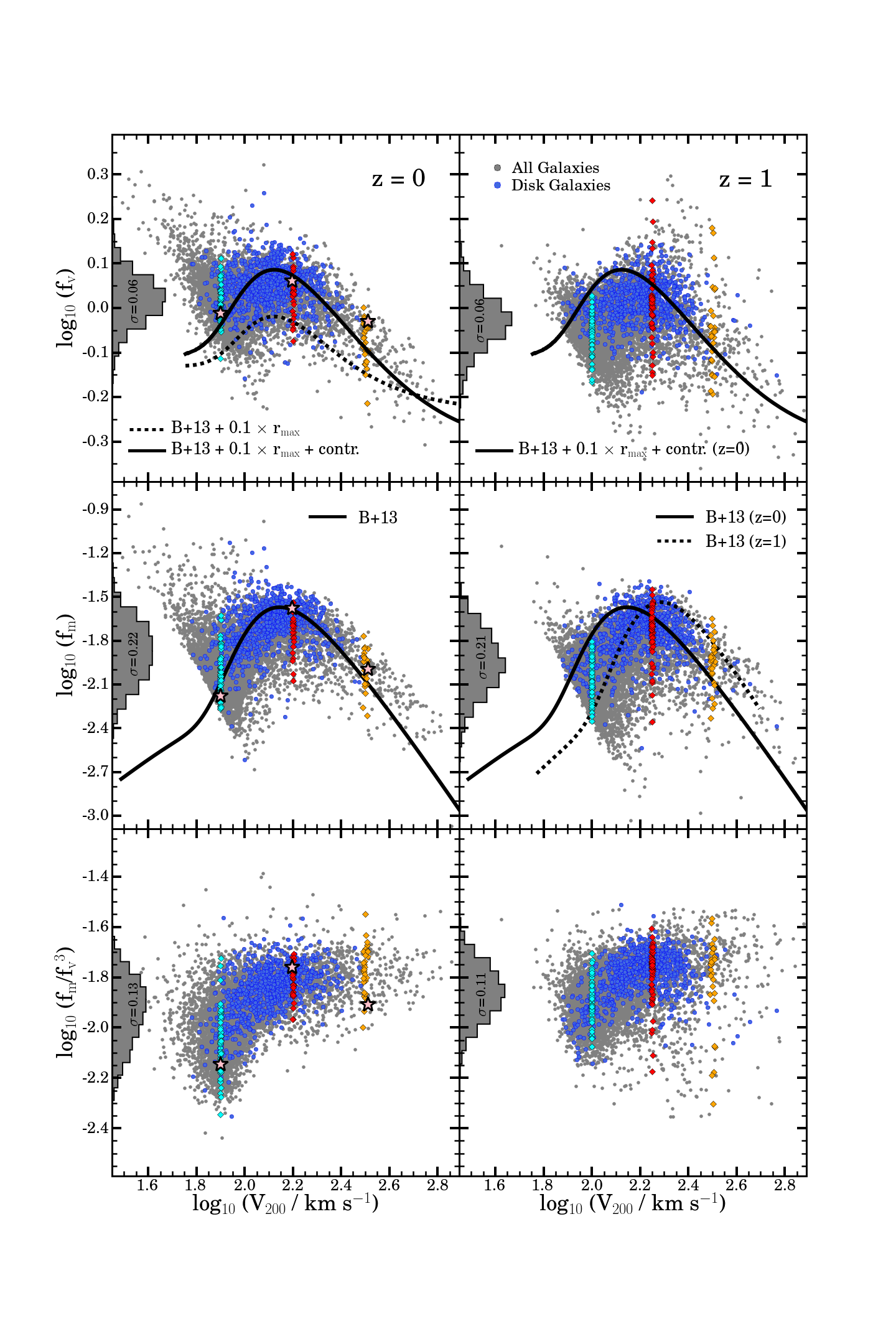} 
\vspace*{-20mm}
\caption{From top to bottom, the rows show respectively the velocity
  ratio parameter $f_v \equiv V_{\rm rot}/V_{200}$; the galaxy
  formation efficiency parameter, $f_m \equiv M_{\rm str}/M_{200}$;
  and the ratio $f_m/f_v^3$, as a function of virial velocity at $z=0$
  (left) and at $z=1$ (right). Colors and symbols are as in
  Fig.~\ref{FigSimTFR}. Solid curves correspond to the same model
  discussed in Fig.~\ref{FigModel}. The dashed curve in the $f_v$
  panel is the ``no contraction'' case of Fig.~\ref{FigModel}.}
\label{FigFmFv}
\end{figure*}
\end{center}

\begin{center} 
\begin{figure*}    
\advance\leftskip-0.5cm
\includegraphics[width=180mm]{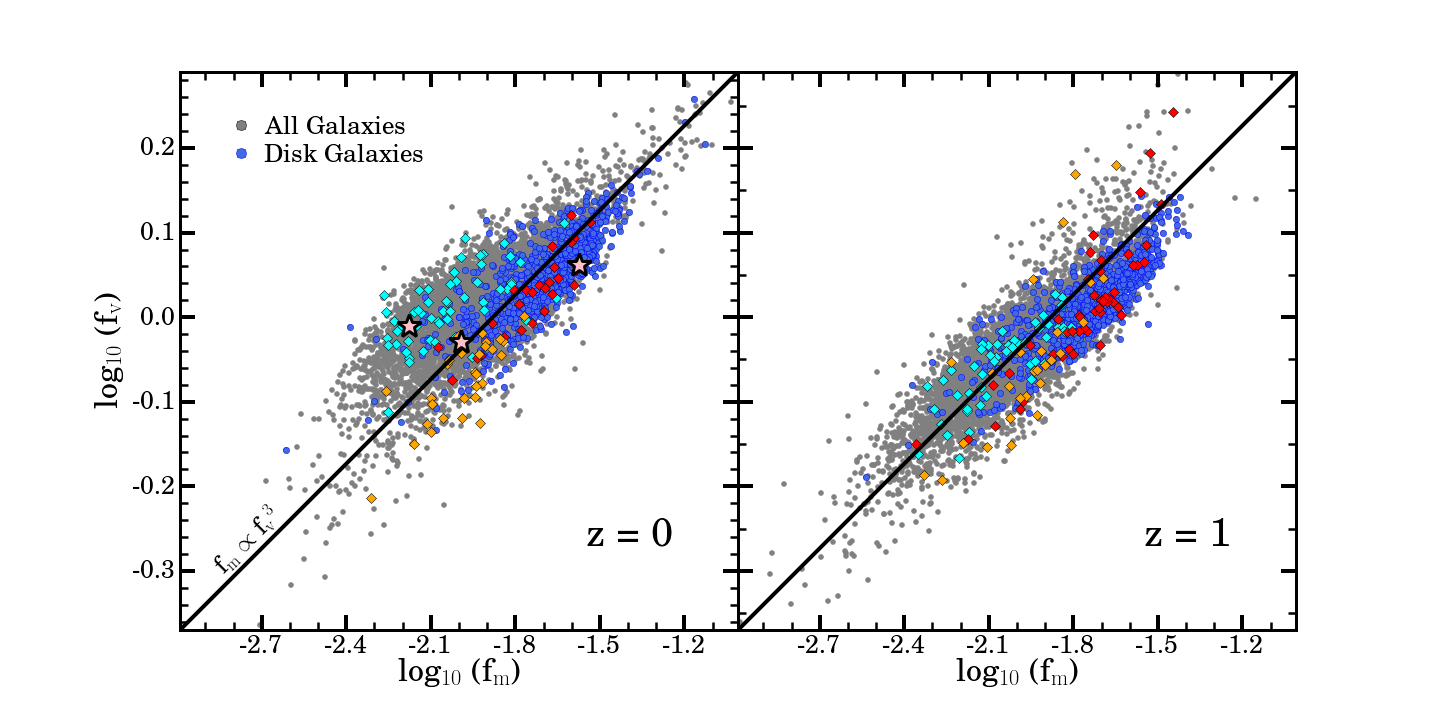} 
\vspace*{-3mm}
\caption{Correlation between the galaxy formation efficiency parameter
  $f_m \equiv M_{\rm str}/M_{200}$, and the velocity ratio parameter,
  $f_v \equiv V_{\rm rot}/V_{200}$, at $z=0$ (left) and at $z=1$
  (right). Colours and symbols are as in Fig.~\ref{FigSimTFR}. The
  relation $f_m\propto f_v^3$ is shown for reference as a straight
  solid line.}
\label{FigScatter}
\end{figure*}
\end{center}

\vspace*{-10mm}
\subsection{TFR slope, zero-point, and scatter}
\label{SecScatter}

The Tully-Fisher relation in $\Lambda$CDM is governed by the relation
between galaxy formation efficiency, $f_m$, and the velocity ratio
parameter, $f_v$, defined in Sec.~\ref{SecModel}. This is clear from
Eq.~\ref{EqM200V200}, which may be re-written as
\begin{equation}
M_{\rm str}=f_mM_{200}={f_m V_{200}^3\over 10GH(z)}=
{1 \over 10GH(z)} {f_m\over f_v^3} V_{\rm rot}^3.
\label{EqFmFv}
\end{equation}
The TFR is thus largely set by the ratio $f_m/f_v^3$: (i) its mass
dependence determines the TFR slope; (ii) its actual value at some
reference mass fixes the TFR zero-point; and (iii) its dispersion
controls the TFR scatter. We explore these issues in more detail in
Fig.~\ref{FigFmFv}, where the left panels show $f_m$, $f_v$, and $f_m/f_v^3$, as
a function of halo virial velocity for EAGLE galaxies at $z=0$.

Fig.~\ref{FigFmFv} shows that, although $f_m$ and $f_v$ have a complex
dependence on virial velocity, they are strongly correlated, resulting
in a $f_m/f_v^3$ ratio that increases monotonically with halo
mass. The monotonic trend ``straightens'' the resulting TFR into a
single power law that matches observations. 

The correlation between $f_m$ and $f_v$ is heavily dependent on galaxy
sizes. Consider the solid line in the left $f_v$ panel of
Fig.~\ref{FigFmFv}, which show the result of applying the simple
adiabatic contraction model of Fig.~\ref{FigModel} to galaxies that
satisfy the B+13 AM prescription, as well as the mass-size relation of
the P+07 sample. The combination implies a non-monotonic mass
dependence for $f_v$, resembling that of $f_m$. The magnitude of the
effect on $f_v$ depends on the actual sizes of the galaxies and on
halo response. The ``no contraction'' case is shown with a dashed
curve.  As discussed in Sec.~\ref{SecModel}, galaxy size and halo
contraction play a crucial role in straightening the TFR into a power
law.

We examine next the TFR zero-point by choosing, as reference, halos
with $V_{200}=160$ km/s (or $2.2$ in $\log_{10}$ units).  For such
halos, on average, $f_m=0.02$ (set by AM) and $f_v=1.08$ (set by size
plus contraction); this implies
$M_{str}=2.7\times10^{10}\, $M$_{\odot}$ at
$V_{\rm rot}=f_v\, V_{200}=171$ km/s, in excellent agreement with
observations, as judged from Fig.~\ref{FigSimTFR2}.  

Finally, Eq.~\ref{EqFmFv} shows that the TFR scatter depends on the
dispersion in $f_m/f_v^3$ rather than on that in $f_m$ or $f_v$,
independently. Indeed, as discussed by \citet{Navarro2000}, the
surprisingly small scatter in the simulated TFR results from the
strong mass-velocity correlation linking galaxies that form in halos
of the same virial mass.

This is clearly illustrated by the colored dots in the right-hand
panel of Fig.~\ref{FigSimTFR} which show that galaxies formed in halos
of fixed virial mass scatter along the main relation, compensating
variations in galaxy formation efficiency with correlated changes in
velocity.  Such a correlation between $f_m$ and $f_v$ is expected,
since, other things being equal, a disk mass increase will
generally lead to a larger circular velocity. Quantitatively, the effect
depends on the gravitational importance of the disk relative to that of the
dark matter. Expressing this in terms of $\nu_{\rm str} \equiv V_{\rm
  str}/V_{\rm dm}$; i.e., the ratio between the stellar and dark
matter contributions to the circular velocity at $r_h$, $V_{\rm
  rot}=(V^2_{\rm str}+V^2_{\rm dm})^{1/2}$, we can write
\begin{equation}
  \delta \ln V_{\rm rot}={1\over 2} {\nu_{\rm str}^2\over (1+\nu_{\rm
      str}^2)} \delta \ln M_{\rm str}.
\end{equation}
A change in galaxy mass, $M_{\rm str}$, then induces a correlated
change in velocity that is stronger the more important the disk
is. For disks that contribute half of the total mass within the
stellar half-mass radius ($\nu_{\rm str}=1$), we would then expect
$f_m\propto f_v^4$, and $f_m\propto f_v^3$ for systems as
baryon-dominated as $\nu_{\rm str}=\sqrt 2$. Although these trends
neglect the effect of contraction, they account for the simulation
results quite well, as may be seen from the $f_m$-$f_v$ correlation
shown in Fig.~\ref{FigScatter}.

In other words, if disks are gravitationally important, then the TFR
scatter is expected to be lower than the scatter in $f_m$ or $f_v$
alone.  Baryons are indeed relatively important in EAGLE galaxies: at
$z=0$ their contribution increases steadily with mass/velocity,
reaching about half of the mass within $r_h$ for $V_{\rm rot} = 150$
km/s. This leads to the strong correlation between $f_m$ and $f_v$
shown in Fig.~\ref{FigScatter} that drastically limits the scatter in the
TFR: although the rms of $f_m$ and $f_v$ are $0.22$ dex and $0.06$
dex, respectively, that of the ratio $f_m/f_v^3$ is just $0.13$ dex.

\begin{center} 
\begin{figure*}  
\advance\leftskip-0.4cm
\includegraphics[width=190mm]{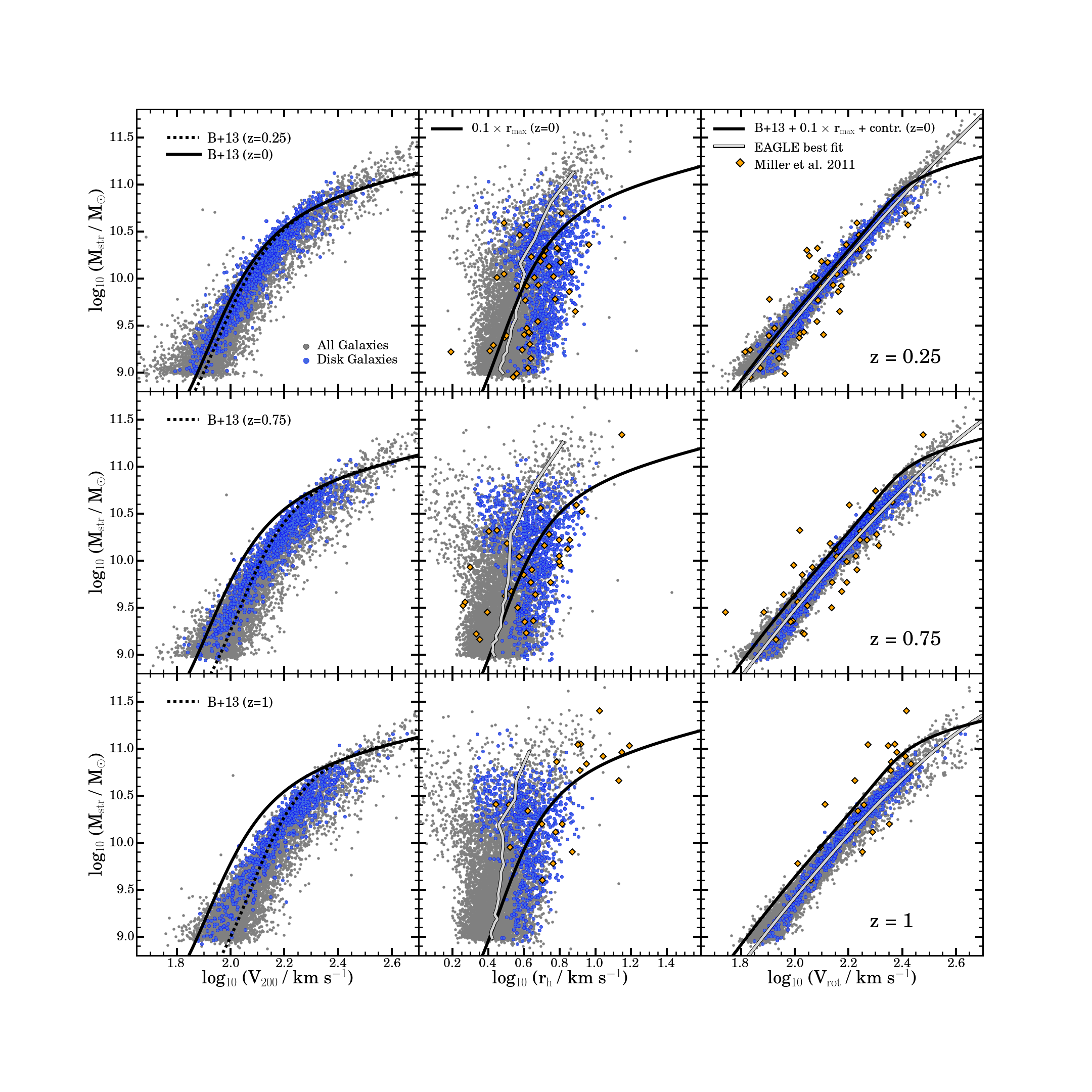} 
\vspace*{-20mm}
\caption{Evolution of the simulated Tully-Fisher relation,
  compared with data from \citet{Miller2011} (orange symbols). Each
  row is as in Fig.~\ref{FigSimTFR}. Solid black lines in each panel
  are the same as in Fig.~\ref{FigModel}, and are included to aid
  comparison. The dotted curves in the left panels indicate the
  predictions from the B+13 abundance-matching model. The thin grey
  lines are fits to EAGLE galaxies, with parameters given in
  Table~\ref{TabFitParam}. EAGLE galaxies match the AM predictions reasonably
  well, as well as the size and TFR of galaxies in
  the observed sample at all redshifts. See text for a more in-depth
  discussion. }
\label{FigTFEvol}
\end{figure*}
\end{center}

\subsection{The evolution of the simulated TFR}
\label{SecTFEVol}

According to the above discussion, the evolution of the TFR with
redshift will depend critically on how galaxy formation efficiencies
and sizes evolve with redshift. We present our results in
Fig.~\ref{FigTFEvol} for three different redshifts: $z=0.25$ (top row), 
$z=0.75$ (middle row) and $z=1$ (bottom row). Each row presents, as 
in Fig.~\ref{FigModel}, the galaxy-halo mass relation on the left, 
the galaxy mass-size relation in the middle, and the TFR in 
the rightmost panel. To guide the comparison, the solid black curves
in each panel are the same as those shown in Fig.~\ref{FigModel} for $z=0$.

From top to bottom, the left-hand panels show that, at given
$V_{200}$, EAGLE galaxy masses decrease with increasing redshift. This
steady decrease in galaxy formation efficiency matches well that
predicted by AM models, as shown by the dotted curves in each panel,
taken from B+13. This is consistent with the agreement between the evolution
of the simulated and observed galaxy mass function found by \cite{Furlong2015a}.
 Note that the evolution in stellar mass is expected
to be especially rapid at the low-mass end of the range studied
here. At $V_{200}=100$ km/s, for example, the stellar mass drops,
according to B+13, by nearly an order of magnitude from $z=0$ to
$z=1$.

Simulated galaxy sizes also evolve, as shown in the middle panels of
Fig.~\ref{FigTFEvol}. The evolution is especially noticeable at the
massive end, dropping by roughly a factor of two back to $z=1$ from
present values. The evolution is weaker, but still noticeable, at the
low-mass end. Interestingly, the half-mass radii of simulated galaxies
at early times agrees fairly well with those in the Tully-Fisher
samples of \citet{Miller2011}. The only difference is at $z=1$, where
at the high-mass end the observed galaxies seem significantly larger
than simulated galaxies of similar mass. The good match with
observations is consistent with the results of \cite{Furlong2015b}, who found that
EAGLE reproduces the observed size evolution relatively well.

The TFR at each redshift is shown in the right-hand column of
Fig.~\ref{FigTFEvol}, and shows good agreement with the data of
\citet{Miller2011}. The evolution in size partly compensates for the
decrease in galaxy formation efficiency at fixed virial velocity, as
shown in the $z=1$ panels of Fig.~\ref{FigFmFv}. This shifts galaxies
closer to the $z=0$ relation, weakening the evolution of the resulting
TFR.

\begin{table}
\begin{center}
\advance\leftskip-0.5cm
\caption{\label{TabFitParam} EAGLE best TFR fit parameters.}
\vspace*{-5mm}
\resizebox{9cm}{!} {
\begin{tabular}{@{}c|c|c|c|c|c|l@{}}
\toprule
Redshift  & \textbf{$M_0 [\rm \times 10^{8}]$}   & \textbf{$\alpha$}  & \textbf{$\gamma$}  & \textbf{$\sigma_{M} [\rm dex]$}   & \textbf{$\sigma_{V}[\rm dex]$}   &  \\ \midrule
$0.00$    & $8.63\pm0.11$     			 & $4.10\pm0.05$      & $0.432\pm0.021$    & $0.11$                 	       & $0.08$                 	  &  \\ 
$0.25$    & $8.32\pm0.12$    			 & $4.13\pm0.04$      & $0.483\pm0.016$    & $0.10$                  	       & $0.07$                 	  &  \\ 
$0.75$    & $6.60\pm0.07$    			 & $4.36\pm0.03$      & $0.594\pm0.009$    & $0.09$                 	       & $0.08$                 	  &  \\
$1.00$    & $5.57\pm0.06$    			 & $4.46\pm0.03$      & $0.634\pm0.008$    & $0.09$                  	       & $0.08$                 	  &  \\
\hline
\multicolumn{6}{l}{\textit{$M_{\rm str}/{\rm M}_\odot=M_0 \,
    \nu^{\alpha} \exp(-\nu^{\gamma})$, \rm where $\nu$ is the  velocity in units of $50$ km/s.}} &  \\
\bottomrule

\end{tabular} 
}
\end{center}

\end{table}

The exception is at $z=1$, where observed galaxies have slightly lower
velocities than in the simulation. Recalling the discussion in
Sec.~\ref{SecModel}, the reason for the offset is most likely driven
by the mismatch in galaxy sizes. At given mass, the larger the galaxy
the lower the contribution of the disk to the circular velocity and
the weaker the halo response, leading, on average, to lower
velocities. Had our simulation produced galaxies as massive and large
as those in the $z=1$ \citet{Miller2011} sample, it is quite likely
that they would have matched the observed velocities.

It is unclear at this point whether the lack of large, massive disks at $z=1$
in EAGLE is a problem for the model or a result of the relatively
small simulated volume, coupled with inherent selection biases
affecting observational samples. Indeed, no such discrepancy was found by
\cite{Furlong2015b}, who compared EAGLE with data from \cite{vanderWel2014}. 
Large, massive disks are obviously
easier to observe at high redshift: given the sensitivity of the TFR
to galaxy size, this has the potential of inducing biases in the
relation that ought to be carefully taken into account. With this
caveat, we conclude that the overall TFR evolution seen in EAGLE is
quite consistent with presently available observational constraints.

\section{Summary and Conclusions}
\label{SecConc}

We have used the EAGLE set of $\Lambda$CDM cosmological hydrodynamical
simulations to study the relation between abundance matching, galaxy
sizes, and the Tully-Fisher relation (TFR). Our main findings may be
summarized as follows:

\begin{itemize}

\item Galaxies that match the predictions of abundance
  matching are consistent with the observed TFR despite the
  non-monotonic behaviour of the galaxy formation efficiency with halo
  mass.  Consistency with the observed TFR requires galaxies to follow
  the mass-size relation of observed galaxy disks, and halos to
  respond to galaxy assembly by contracting, roughly as predicted by
  simple adiabatic contraction models. 

\item EAGLE galaxies match all of these constraints approximately, and
  show a Tully-Fisher relation in good agreement with observation at
  $z=0$.

\item Galaxy size and halo contraction induce a strong correlation
  between galaxy formation efficiency and rotation speed that
  straightens the TFR into a single power law and scatters galaxies
  along the main relation, keeping its dispersion tight. As a result,
  the scatter of the simulated TFR is substantially lower than
  observed, even when {\it all} galaxies are considered, rather than
  only late-type disks.

\item The EAGLE galaxy-halo mass relation evolves 
  roughly as expected from AM models and its galaxy mass-size relation
  evolves in accord with that of galaxies in distant Tully-Fisher
  samples. This results in gradual but weak evolution of the simulated
  TFR in reasonable agreement with observed constraints, despite the more
  rapid evolution in galaxy formation efficiency predicted by
  abundance-matching models.

\end{itemize}

The sensitivity of the Tully-Fisher relation to galaxy size explains
why many simulations have argued consistency with this scaling
relation while, at the same time, failing to match the galaxy
masses predicted by abundance-matching models. Indeed,
it is always possible to trade disk mass for galaxy size so that the
resulting TFR remains nearly invariant. A galaxy formation simulation
cannot therefore be considered successful unless it matches
simultaneously the Tully-Fisher relation, as well as the abundance and
size of galaxies as a function of stellar mass. Overall, our results
show that the slope, zero-point, scatter, and evolution of the TFR
pose no obvious difficulty to galaxy formation models in the
$\Lambda$CDM cosmogony.

\section*{Acknowledgments}
\label{SecAck}

IF, MGA, JFN, LVS acknowledge financial support of grant
PICT-1137/2012 from the Agencia Nacional de Promoci\'on Cient\'ifica y
Tecnol\'ogica, Argentina. IF, MGA acknowledge financial support of
grant 203/14 of the SECYT-UNC, Argentina. The research was supported
in part by the European Research Council under the European Union\'s
Seventh Framework Programme (FP7/2007-2013) / ERC Grant agreement
278594-GasAroundGalaxies, 267291-COSMIWAY and by the Interuniversity
Attraction Poles Programme initiated by the Belgian Science Policy
Office (AP P7/08 CHARM). This work used the
DiRAC Data Centric system at Durham University, operated by the
Institute for Computational Cosmology on behalf of the STFC DiRAC HPC
Facility (www.dirac.ac.uk). This equipment was funded by BIS National
E-infrastructure capital grant ST/K00042X/1, STFC capital grants
ST/H008519/1 and ST/K00087X/1, STFC DiRAC Operations grant
ST/K003267/1 and Durham University.  DiRAC is part of the National
E-Infrastructure. RAC is a Royal Society University Research Fellow.

\bibliographystyle{mn2e}
\bibliography{references}

\end{document}